\documentclass[conference]{IEEEtran}
\IEEEoverridecommandlockouts
\usepackage{cite}
\usepackage{amsmath,amssymb,amsfonts}
\usepackage{algorithmic}
\usepackage{graphicx}
\usepackage{adjustbox}
\usepackage{lscape}
\usepackage{textcomp}
\usepackage{xcolor}
\usepackage{listings}
\usepackage{tcolorbox} 
\usepackage{tabularx}
\usepackage{array}
\usepackage{booktabs} 
\usepackage{multirow} 
\usepackage{makecell}
\usepackage{hyperref}
\usepackage{float}
\usepackage{amsmath}

\definecolor{red}{HTML}{de2d26}
\definecolor{green}{HTML}{31a354}

\begin{document}

\title{Advanced U-Net Architectures with CNN Backbones for Automated Lung Cancer Detection and Segmentation in Chest CT Images\\
}

\author{
\IEEEauthorblockN{1\textsuperscript{st} Alireza Golkarieh}
\IEEEauthorblockA{
PhD student in Computer Science and Informatics\\
Oakland University, Rochester, Michigan, USA\\
Email: golkarieh@oakland.edu}
\and
\IEEEauthorblockN{2\textsuperscript{nd} Kiana Kiashemshaki}
\IEEEauthorblockA{
Master of Science in Computer Science\\
Department of Computer Science\\
Bowling Green State University, Ohio, USA\\
Email: kkiana@bgsu.edu}
\and
\IEEEauthorblockN{3\textsuperscript{rd} Sajjad Rezvani Boroujeni}
\IEEEauthorblockA{
Master of Science in Applied Statistics\\
Department of Applied Statistics \& Operations Research\\
Bowling Green State University, Ohio, USA\\
Email: sajjadr@bgsu.edu}
\and
\IEEEauthorblockN{4\textsuperscript{th} Nasibeh Asadi Isakan}
\IEEEauthorblockA{
PhD student, Biomedical Engineering\\
College of Engineering\\
University of Kentucky, Lexington, Kentucky, USA\\
Email: nas237@uky.edu}
}

\maketitle

\begin{abstract}
Objective: This study aims to evaluate the effectiveness of U-Net architectures integrated with different convolutional neural network (CNN) backbones for automated lung cancer detection and segmentation in chest CT images, addressing the critical need for accurate and efficient diagnostic tools in clinical practice.
Methods: A balanced dataset of 832 chest CT images (416 cancerous and 416 non-cancerous cases) was preprocessed using Contrast Limited Adaptive Histogram Equalization (CLAHE) and resized to 128×128 pixels. U-Net models with three distinct CNN backbones (ResNet50, VGG16, and Xception) were implemented for lung region segmentation. Subsequently, CNN classifiers trained on segmented images and hybrid approaches combining CNN feature extraction with machine learning algorithms (Support Vector Machine, Random Forest, and Gradient Boosting) were evaluated. Performance was assessed using 5-fold cross-validation with metrics including accuracy, precision, recall, F1-score, Dice coefficient, and ROC-AUC.
Results: For segmentation tasks, U-Net with ResNet50 achieved superior performance in cancerous lung detection with a Dice score of 0.9495 ± 0.1089 and accuracy of 0.9735 ± 0.1015, while U-Net with VGG16 excelled in non-cancerous segmentation (Dice: 0.9532 ± 0.0993, accuracy: 0.9513 ± 0.1006). In classification, the CNN model with U-Net-Xception backbone demonstrated exceptional performance with 99.1\% accuracy, 99.74\% recall, and 99.42\% F1-score. Hybrid approaches yielded competitive results, with CNN-SVM-Xception achieving 96.7\% accuracy and 97.88\% F1-score. Comparative analysis revealed that our methodology outperformed several recent studies in both segmentation and classification metrics.
Conclusion: The integration of U-Net architectures with advanced CNN backbones provides a robust framework for lung cancer detection, demonstrating superior performance in both segmentation and classification tasks. The synergistic relationship between accurate segmentation and enhanced classification accuracy establishes a reliable foundation for clinical implementation. This approach offers significant potential for improving diagnostic precision, reducing radiologist workload, and facilitating early lung cancer detection in clinical settings, ultimately contributing to better patient outcomes and healthcare efficiency.

\end{abstract}


\section{Introduction}

Chest cancer, which includes malignancies of the lungs, esophagus, and chest wall, poses a significant public health challenge due to its high incidence and associated mortality rates. Among these, lung cancer is particularly prominent, consistently identified as the leading cause of cancer-related deaths globally. Data from the World Health Organization (WHO) indicates that in 2020 alone, lung cancer accounted for approximately 2.21 million new cases and 1.8 million deaths worldwide. Other thoracic cancers, such as esophageal cancer, also contribute substantially to the overall impact of chest malignancies \cite{b1}.
It is very important that cancers of the chest be treated as soon as possible. Most of these cancers are usually diagnosed at a later stage, where treatment modalities are limited; thus, survival rates also decrease considerably. Advanced imaging diagnostic tools, such as low-dose CT scans, have been very instrumental in diagnosing cancers of the chest when they are at more salvageable stages. Active research and innovation also help in fighting chest cancer. The efforts should be directed toward better screening programs, new therapeutic options, and awareness of important risk factors like tobacco exposure, environmental exposures, and genetic predispositions. It would further open up opportunities for targeted and more effective therapeutic options if the molecular and genetic basis of these malignancies is understood\cite{b2}.
The detection and classification of lung cancer using CT scans have been a focus of extensive research, with various deep learning models achieving notable success. Shen et al. \cite{b3}, for example, reported that a deep learning-based method significantly enhanced lung nodule detection and classification, reaching an accuracy of 86.84\% through a multi-view CNN architecture. Similarly, Liao et al. \cite{b4} applied transfer learning with a pre-trained ResNet model, achieving 88.5\% accuracy in classifying lung cancer from CT images. Ardila et al. \cite{b5} developed a deep learning algorithm capable of predicting lung cancer risk from a single low-dose CT scan, showing 94.4\% sensitivity and 81.0\% specificity, comparable to expert radiologists. Nam et al. \cite{b6} further achieved an accuracy of 89.6\% using a deep residual learning framework to differentiate malignant from benign nodules. Recent research has continued to explore new methods in lung disease segmentation and classification. Michael Osadebey et al. applied a three-stage segmentation approach combining U-Net and CNN, obtaining Dice scores between 0.76 and 0.95 on the 3DIRCAD and ILD datasets \cite{b7}. In another study, Prachaya Khomduean (2023) introduced a 3D U-Net model combined with DenseNet and ResNet to analyze COVID-19 patient data, achieving a Dice score of 91.52\% for lung lobes and 76.89\% for lesion segmentation, with an R² value of 0.842 \cite{b8}. Likewise, Surbhi Vijh (2023) attained a 97.18\% accuracy in lung nodule classification using a combination of CNN and LDA on 120 CT scans, reporting sensitivity and specificity of 97\% and 98.66\%, respectively \cite{b9}. Moreover, a review by Ting-Wei Wang (2024) analyzed 20 studies comparing deep learning models to radiologists, finding an average sensitivity of 82\% and specificity of 75\% across the studies \cite{b10}. In another example, Vijay Kumar Gugulothu (2024) utilized the LNDC-HDL and HDE-NN techniques for lung nodule classification, achieving 96.39\% accuracy, 95.25\% sensitivity, and an AUC of 96.05\%. Additionally, S.R. Vijayakumar (2024) leveraged CapsNet and U-Net to report a classification accuracy of 98\%, with a precision rate of 97.9\% and a false positive rate of 1.9\% \cite{b11}. For segmentation tasks, Murat Canayaz (2024) employed a combination of EffxNet, InceptionV3, and DenseNet121, achieving a Dice coefficient of 0.8877 and a classification accuracy of 97.98\% \cite{b12}. Similarly, S. Akila Agnes (2024) used Wavelet U-Net++ on the LIDC-IDRI dataset, yielding a Dice coefficient of 0.936 and an IoU of 0.878 \cite{b13}. Finally, Wei Chen (2023) developed a Hybrid Segmentation Network (HSN) specifically for small-cell lung cancer (SCLC), achieving high Dice scores alongside 96.05\% sensitivity and 96.39\% specificity \cite{b14}.
The primary objective of this study is to address the existing gaps in the segmentation and classification of lung diseases by proposing a novel hybrid framework that integrates morphological processing, U-Net models with diverse backbones (VGG16, ResNet50, Xception), and advanced machine learning classifiers (Random Forest, Gradient Boosting, and SVM). Unlike previous studies that predominantly focused on either segmentation or classification tasks independently, this research emphasizes a comprehensive approach that combines robust segmentation techniques with precise classification methodologies. To enhance segmentation accuracy, this study incorporates morphological preprocessing to refine the lung region and leverages U-Net architectures optimized with varying backbone networks to adapt to the diverse anatomical structures present in CT scans. Furthermore, the integration of deep learning-derived features with machine learning classifiers ensures a high level of adaptability for distinguishing cancerous from non-cancerous regions. The novelty of this study lies in its multi-faceted hybrid strategy, which not only bridges the gap between segmentation and classification but also achieves high performance across key metrics such as accuracy, sensitivity, and specificity. By evaluating these models across multiple performance dimensions, this work contributes significantly to the early detection and precise classification of chest cancer, offering a potential pathway for improved clinical decision-making and patient outcomes.

\section{Material and Methods }
\subsection{Data Collection}
In this study, a publicly available dataset was utilized for the analysis of chest CT scans related to lung cancer. The dataset comprises images acquired from the National Institute of Diseases of the Chest and Hospital (NIDCH) in Dhaka, Bangladesh. It originally contained 1050 malignant (cancerous) and 416 normal (healthy) images. To address class imbalance and ensure fair comparison between diagnostic groups, a subset of 416 malignant images was randomly selected to match the number of normal cases, resulting in a balanced dataset of 832 CT images. All images were annotated under the clinical supervision of Dr. Sabina Akter, MBBS, MD (Radiology \& Imaging), Associate Professor and Head of the Department of Radiology and Imaging at NIDCH. The labeling process followed clinical guidelines to ensure reliability and consistency, making the dataset suitable for scientific research purposes. Furthermore, all images were resized to a standard resolution of 512×512 pixels to maintain uniformity in data dimensions. A summary of the number of images per diagnostic category, along with the applied image resolution, is provided in~\ref{table1}.

\begin{table}[h!]
\centering
\caption{Distribution of CT Images by Class After Balancing}
\label{table1}
\begin{tabular}{|l|c|c|}
\hline
\textbf{Diagnostic Category} & \textbf{Number of Images} & \textbf{Image Resolution} \\
\hline
\textbf{Malignant (Cancer)} & 416 & 512 × 512 pixels \\
\textbf{Normal (Healthy)} & 416 & 512 × 512 pixels \\
\hline
\textbf{Total} & 832 & 512 × 512 pixels \\
\hline
\end{tabular}
\end{table}

\subsection{Preprocessing of CT Scan Data}
Preprocessing is an essential phase to ensure the CT images are ready for accurate and efficient classification. The following steps were performed:
\subsubsection{Local Contrast Enhancement with CLAHE}
CLAHE (Contrast Limited Adaptive Histogram Equalization) was used to enhance the local contrast in the images. This method divides the image into smaller sections and applies histogram equalization, improving contrast without amplifying noise.

\begin{equation}
I' = \text{CLAHE}(I)
\label{eq:clahe}
\end{equation}

The parameters used include a \textit{clipLimit} of 2.0 and a \textit{tileGridSize} of (8, 8).

\subsubsection{Uniform Image Rescaling}
The images were resized to a uniform dimension of 224 × 224 pixels to ensure consistency in input size for the neural network models. 
\begin{equation}
I'(x, y) = I\left( \frac{x \cdot \text{width}}{128}, \frac{y \cdot \text{height}}{128} \right)
\label{eq:resize}
\end{equation}

\subsubsection{Pixel Value Normalization}
To facilitate faster convergence during model training, the pixel values were normalized, scaling them to fall between 0 and 1.

\begin{equation}
I' = \frac{I}{255.0}
\label{eq:normalize}
\end{equation}

\section{Automatic Lung Segmentation Using Morphological Techniques}

This section outlines the lung segmentation process, which employs morphological operations like binarization, labeling, and masking. The key steps include:

\textbf{Step 1: Thresholding with Otsu’s Method:}

Otsu’s method is utilized to separate the lung regions from the background by determining an optimal threshold $T_{\text{otsu}}$, minimizing intra-class variance:

\begin{equation}
T_{\text{otsu}} = \arg\min \left( \sigma^2_{\text{B}} + W^2_{\text{B}} \right)
\label{eq:otsu}
\end{equation}

This threshold generates the binary image $B(x, y)$.

\textbf{Step 2: Eliminating Border-Connected Artifacts:}
Border-connected components are removed to eliminate unwanted artifacts:

\begin{equation}
\text{B}_{\text{clear}}(x, y) = \text{B}(x, y) \land \neg\,\text{B}_{\text{border}}(x, y)
\label{eq:border_clear}
\end{equation}

\textbf{Step 3: Expanding Lung Regions with Dilation:}
Dilation is applied using a disk-shaped structuring element (radius 5) to enlarge the lung regions:

\begin{equation}
\text{B}_{\text{dilated}} = \text{B}_{\text{clear}} \oplus \text{D}(5)
\label{eq:dilation}
\end{equation}

\textbf{Step 4: Labeling Connected Components:}
The connected regions are labeled to differentiate individual lung regions:

\begin{equation}
L(x, y) = f\left(\text{B}_{\text{dilated}}(x, y)\right)
\label{eq:labeling}
\end{equation}

\textbf{Step 5: Selecting the Largest Lung Regions:}
The two largest components, representing the lungs, are selected based on their area $A_i$:

\begin{equation}
A_i = \sum_{(x, y) \in R_i} \text{B}_{\text{dilated}}(x, y)
\label{eq:area}
\end{equation}

\textbf{Step 6: Refining Lung Boundaries with Erosion:}
Erosion is performed using a disk (radius 4) to refine the boundaries of the lung regions:

\begin{equation}
\text{B}_{\text{eroded}} = \text{B} \ominus \text{D}(4)
\label{eq:erosion}
\end{equation}

\textbf{Step 7: Closing Gaps with Morphological Closure:}
Morphological closure using a disk (radius 10) is applied to close small gaps within the lung regions:

\begin{equation}
\text{B}_{\text{closed}} = (\text{B}_{\text{eroded}} \oplus \text{D}(10)) \ominus \text{D}(10)
\label{eq:closure}
\end{equation}

\textbf{Step 8: Filling Internal Gaps:}
Any small internal holes in the lung regions are filled:

\begin{equation}
\text{B}_{\text{filled}} = \text{fill}(\text{B}_{\text{closed}})
\label{eq:fill}
\end{equation}

\textbf{Step 9: Applying the Lung Mask:}
The binary mask is superimposed on the original image, preserving only the lung regions:

\begin{equation}
\text{I}_{\text{masked}}(x, y) =
\begin{cases}
\text{I}(x, y), & \text{if } \text{B}_{\text{filled}}(x, y) = 1 \\
0, & \text{otherwise}
\end{cases}
\label{eq:mask}
\end{equation}

This step ensures that only the lung regions are retained, while the background and other structures are set to zero.

\begin{figure*}[!t]
    \centering
    \includegraphics[width=\textwidth]{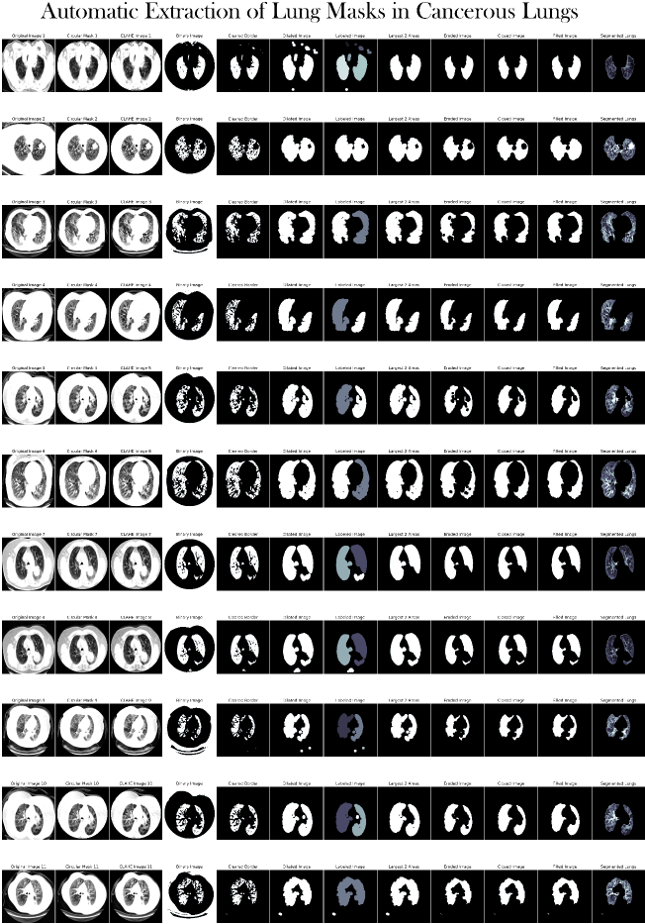}
    \caption{Stepwise Lung Segmentation Process for U-Net Ground Truth Generation}
    \label{fig1}
\end{figure*}

Figure\ref{fig1} illustrates the stepwise lung segmentation process, serving as ground truth for training a U-Net model. Starting from the original lung CT scans, the images undergo several morphological operations: binarization, border clearing, dilation, labeling, selection of the two largest areas, erosion, closing, and hole filling. The final segmented lung masks are then superimposed on the original images.

\section{U-Net Architectures with Pre-trained Backbones for Lung Segmentation}
\subsubsection{U-Net with VGG16 Backbone for Lung Segmentation}

\begin{figure*}[!t]
    \centering
    \includegraphics[width=\textwidth]{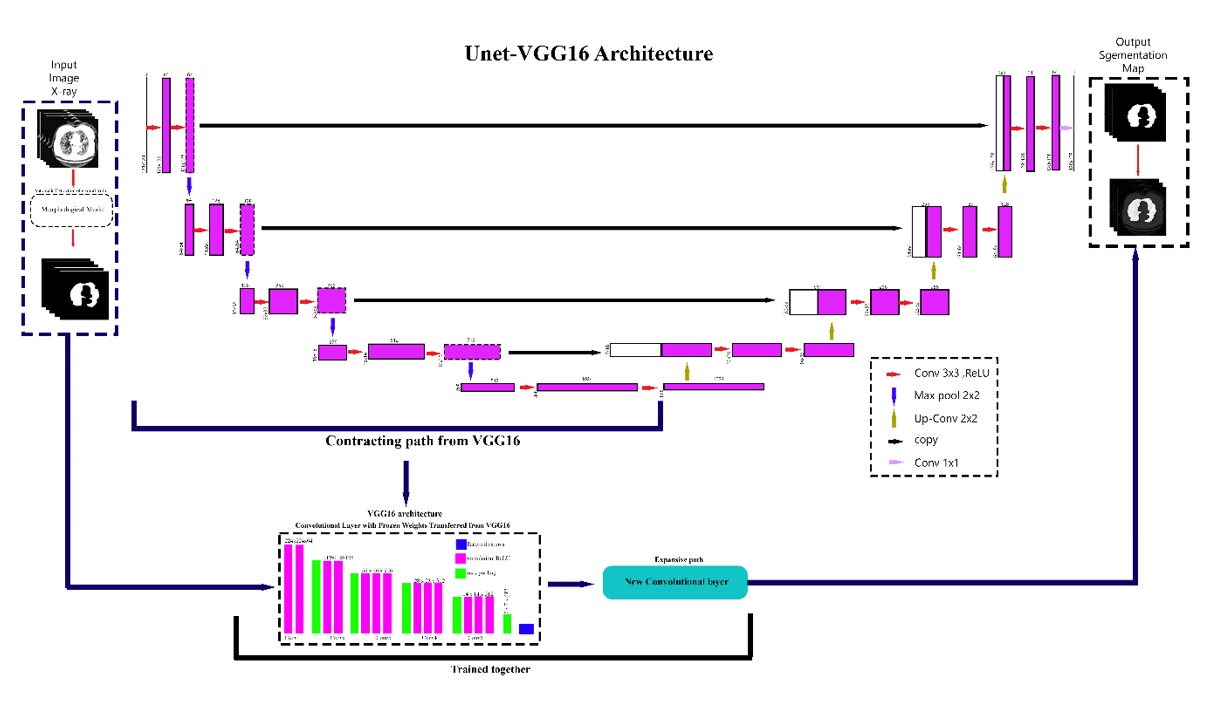}
    \caption{U-Net Architecture with VGG16 Backbone for Lung Segmentation in Chest CT Scan Images}
    \label{fig2}
\end{figure*}

In this research, we utilize a U-Net model integrated with a VGG16 backbone to enhance the accuracy of lung region segmentation in CT scan images. The VGG16 model, pre-trained on the ImageNet dataset, serves as the encoder to extract rich feature representations. The U-Net’s symmetric decoder reconstructs the segmentation map with high precision, utilizing skip connections to retain spatial information \cite{b15}.
The VGG16 backbone, utilized as the encoder in the U-Net architecture, consists of five convolutional blocks, each followed by max-pooling layers to incrementally reduce the spatial dimensions. These blocks produce feature maps that serve as skip connections for the U-Net decoder, ensuring detailed spatial information is preserved. Each convolutional block applies ReLU activations following batch normalization to enhance feature extraction, while the max-pooling layers reduce dimensions. The decoder then reconstructs the segmentation map by using transposed convolutions for upsampling and integrating corresponding encoder features through skip connections. Ultimately, a 1×1 convolution layer with sigmoid activation yields the final segmentation output. The model, compiled with the Adam optimizer and binary cross-entropy loss, is trained with early stopping to mitigate overfitting. Detailed parameters and backbone configurations are provided in Table\ref{table2}. Figure\ref{fig2} depicts the U-Net architecture with a VGG16 backbone used for segmenting lungs in chest CT scan images.

\subsubsection{U-Net with ResNet50 Backbone for Lung Segmentation}
\begin{figure*}[!t]
    \centering
    \includegraphics[width=\textwidth]{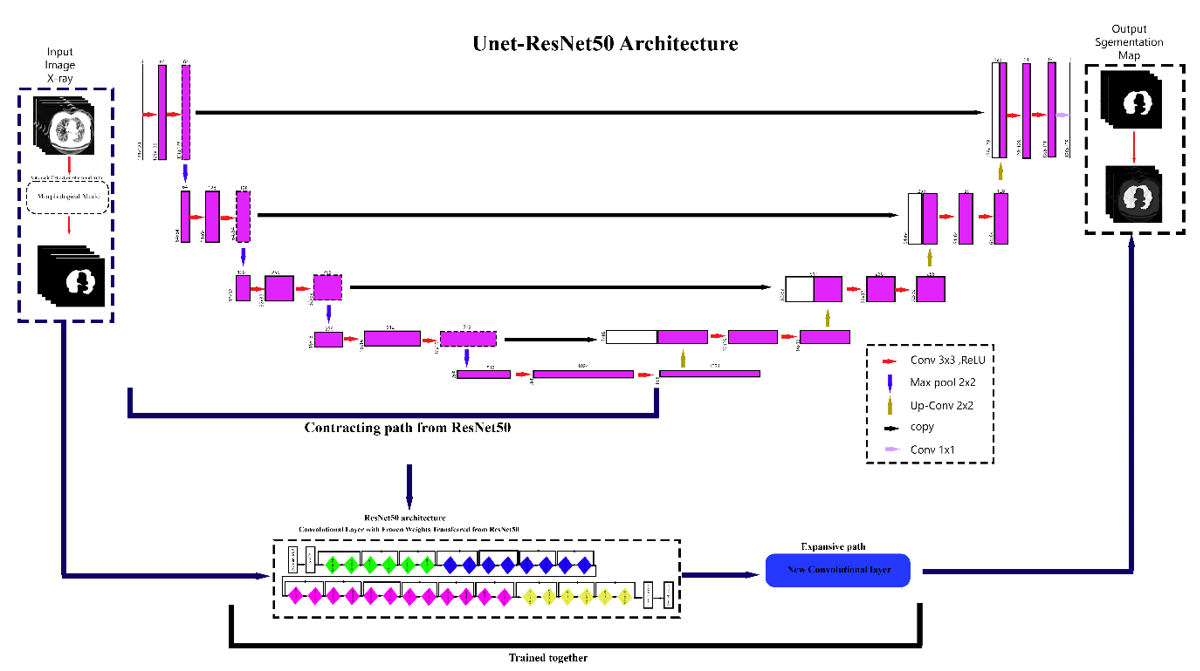}
    \caption{U-Net Architecture with ResNet50 Backbone for Lung Segmentation in Chest CT Scan Images}
    \label{fig3}
\end{figure*}

In this study, we implement a U-Net architecture with a ResNet50 backbone to enhance the segmentation of lung regions in CT scan images. The ResNet50 model, pre-trained on ImageNet, serves as the encoder to extract detailed feature representations. Its structure comprises convolutional blocks with residual connections, which facilitate deep network training by mitigating degradation issues. Specific outputs from ResNet50 blocks are used as skip connections to the U-Net decoder, allowing spatial details to be preserved through the segmentation process [16]. The decoder path, similar to the VGG16-based U-Net, reconstructs the segmentation map with transposed convolutions and concatenated skip connections for accurate feature mapping. Additional model parameters and backbone details are outlined in Table\ref{table2}. Figure\ref{fig3} illustrates the U-Net architecture with a ResNet50 backbone designed for segmenting lung regions in chest CT scan images.

\subsubsection{U-Net with Xception Backbone for Lung Segmentation}
\begin{figure*}[!t]
    \centering
    \includegraphics[width=\textwidth]{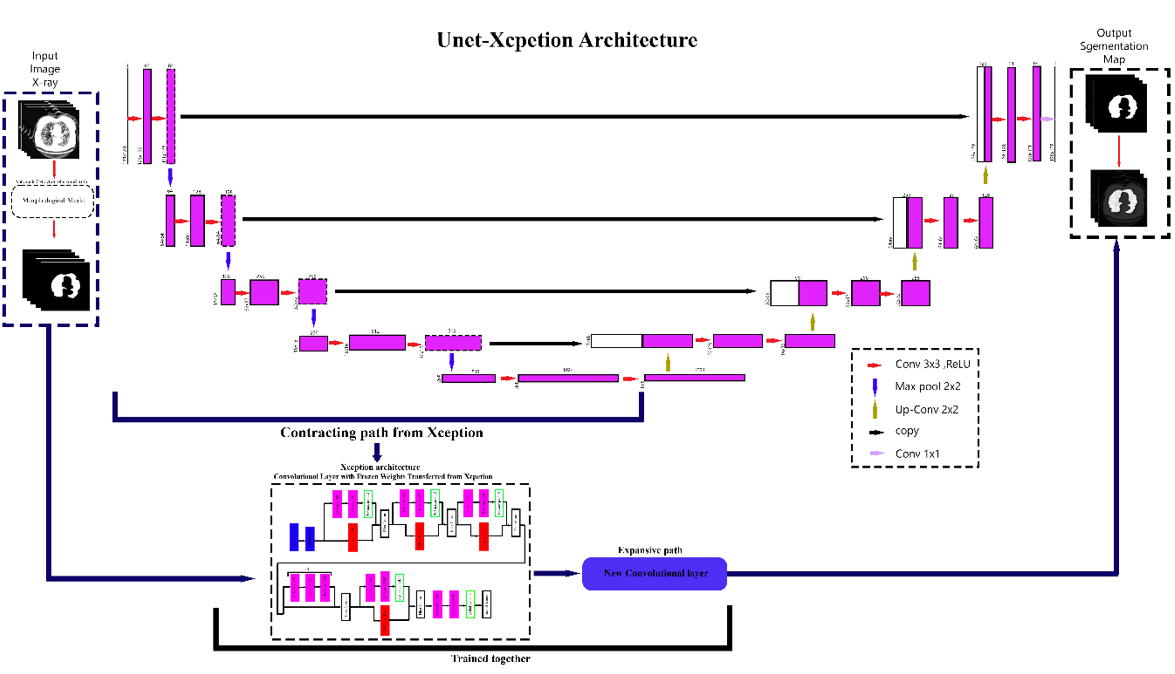}
    \caption{U-Net Architecture with Xception Backbone for Lung Segmentation in Chest CT Scan Images}
    \label{fig4}
\end{figure*}

In this section, we employ a U-Net architecture powered by an Xception backbone to improve the segmentation of lung regions within CT scan images. The Xception model, pre-trained on ImageNet, functions as the encoder to capture comprehensive feature representations. Specific layers from the Xception model provide skip connections to the U-Net decoder, ensuring spatial details are retained throughout the segmentation process \cite{b17}. The decoder uses transposed convolutions for upsampling, with skip connections aligned via cropping to maintain compatibility in feature size. The model configuration closely mirrors the VGG16-based U-Net, as detailed in Table\ref{table2}, to leverage the feature extraction strengths of each backbone. Figure\ref{fig4} presents the U-Net architecture integrated with an Xception backbone, aimed at segmenting lung regions in chest CT scan images.

\begin{table*}[htbp]
\centering
\caption{Architecture and Learning Parameters of U-Net with VGG16, ResNet50, and Xception Backbones for Lung Segmentation}
\label{table2}
\renewcommand{\arraystretch}{1.3}
\begin{tabularx}{\textwidth}{|>{\raggedright\arraybackslash}m{3.5cm}|X|X|X|}
\hline
\textbf{Parameter} & \textbf{VGG16 Backbone} & \textbf{ResNet50 Backbone} & \textbf{Xception Backbone} \\
\hline
Pre-training Dataset & ImageNet & ImageNet & ImageNet \\
\hline
Backbone Layers (Encoder) & 5 Convolutional Blocks, each with Max-Pooling & Residual Blocks with Max-Pooling & Depthwise Separable Convolutions with Max-Pooling \\
\hline
Skip Connection Layers & \makecell[l]{block1\_conv2,\\ block2\_conv2,\\ block3\_conv3,\\ block4\_conv3,\\ block5\_conv3} & 
\makecell[l]{conv1\_relu,\\ conv2\_block3\_out,\\ conv3\_block4\_out,\\ conv4\_block6\_out,\\ conv5\_block3\_out} &
\makecell[l]{block1\_conv1\_act,\\ block3\_sepconv2\_act,\\ block4\_sepconv2\_act,\\ block13\_sepconv2\_act,\\ block14\_sepconv2\_act} \\
\hline
Decoder Path & Transposed Convolutions and Concatenations & Transposed Convolutions and Concatenations & Transposed Convolutions with Cropping and Concatenations \\
\hline
Activation Function & ReLU & ReLU & ReLU \\
\hline
Upsampling Method & Transposed Convolutions & Transposed Convolutions & Transposed Convolutions with Cropping \\
\hline
Output Layer & Conv2D(1, (1,1)) + Sigmoid Activation & Conv2D(1, (1,1)) + Sigmoid Activation & Conv2D(1, (1,1)) + Sigmoid Activation \\
\hline
Loss Function & Binary Cross-Entropy & Binary Cross-Entropy & Binary Cross-Entropy \\
\hline
Optimizer & Adam & Adam & Adam \\
\hline
Early Stopping & Enabled, patience=15 epochs & Enabled, patience=15 epochs & Enabled, patience=15 epochs \\
\hline
Cross-Validation & 5-Fold & 5-Fold & 5-Fold \\
\hline
Data Split & 80\% Training, 20\% Testing/Validation & 80\% Training, 20\% Testing/Validation & 80\% Training, 20\% Testing/Validation \\
\hline
Epochs & 50 & 50 & 50 \\
\hline
Batch Size & 2 & 2 & 2 \\
\hline
\end{tabularx}
\end{table*}

\subsection{Classification using a combination of CNN and machine learning}
\subsubsection{Independent CNN Model}

\begin{figure*}[!t]
    \centering
    \includegraphics[width=\textwidth]{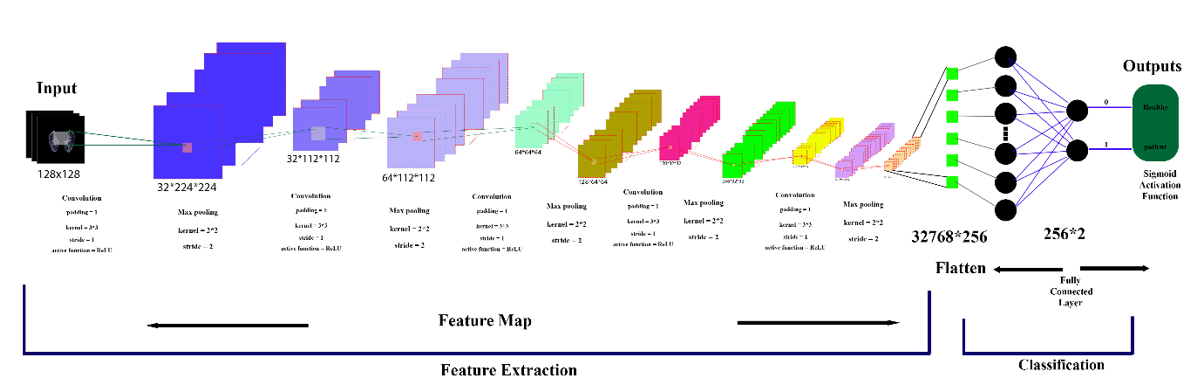}  
    \caption{Architecture of a Convolutional Neural Network (CNN) Model}
    \label{fig5}
\end{figure*}

The CNN model was independently designed to classify chest CT images into cancerous and non-cancerous categories. The model architecture and training process are detailed below:
 Model Architecture
The convolutional neural network (CNN) was specifically designed to classify chest CT images into cancerous and non-cancerous categories. This model leverages a hierarchical structure consisting of convolutional layers for feature extraction, pooling layers for dimensionality reduction, and dropout layers to mitigate overfitting [18]. Key components include Conv2D layers with 3×3 filters and ReLU activation for non-linear feature extraction, MaxPooling layers with a 2×2 window for spatial down-sampling, and dense layers for final classification. The output layer employs a sigmoid activation function to produce probabilities suitable for binary classification. The model is optimized using the Adam optimizer and employs binary cross-entropy as the loss function, ensuring effective optimization for the binary classification task. A batch size of 16, a validation split of 10\%, and early stopping based on validation loss were incorporated to enhance training efficiency and avoid overfitting. The parameters of the CNN, including the use of dropout rates, batch normalization, and regularization techniques, are meticulously designed to improve generalization and robustness. A detailed summary of the model's architecture and hyperparameters is provided in Table 3 for reference. Figure\ref{fig5}, diagrammatic representation of a Convolutional Neural Network (CNN) architecture, illustrating the layers and connections used for feature extraction and classification in deep learning tasks.

\subsubsection{Hybrid Models: CNN with SVM, RF, and GB}
In addition to the standalone CNN model, hybrid models were developed by combining the CNN with traditional machine learning classifiers such as Support Vector Machine (SVM), Random Forest (RF), and Gradient Boosting (GB). These hybrid models leverage the feature extraction capabilities of CNN and the classification strengths of SVM, RF, and GB.
CNN Architecture for Feature Extraction
The CNN architecture used for feature extraction is the same as the standalone model, consisting of convolutional layers, pooling layers, and dropout layers. The output of the final dropout layer is flattened to create feature vectors for input into the traditional classifiers.
Traditional Classifiers for Classification
The flattened feature vectors from the CNN are used as input to three different classifiers: SVM, RF, and GB. Each classifier is trained to predict whether the input image is cancerous or non-cancerous.
Support Vector Machine (SVM): SVM is a supervised learning model that finds the hyperplane that best separates the classes \cite{b19}. 

\begin{equation}
f(x) = \text{sign}(w^{\mathrm{T}} x + b)
\label{eq:svm_decision}
\end{equation}

where w is the weight vector and b is the bias. The SVM model employs an RBF kernel for classification, with various hyperparameters fine-tuned for optimal performance. Table\ref{table3} provides a comprehensive overview of these learning parameters.

Random Forest (RF): RF is an ensemble learning method that uses multiple decision trees to improve classification accuracy \cite{b20}. 

\begin{equation}
f(x) = \frac{1}{T} \sum_{t=1}^{T} h_t(x)
\label{eq:ensemble_avg}
\end{equation}

where T is the number of trees and $h_t$ is the DT classifier. The Random Forest model incorporates 100 estimators to create a robust ensemble classifier. For further details on parameter settings and cross-validation, refer to Table 3.

\begin{figure*}[!t]
    \centering
    \includegraphics[width=\textwidth]{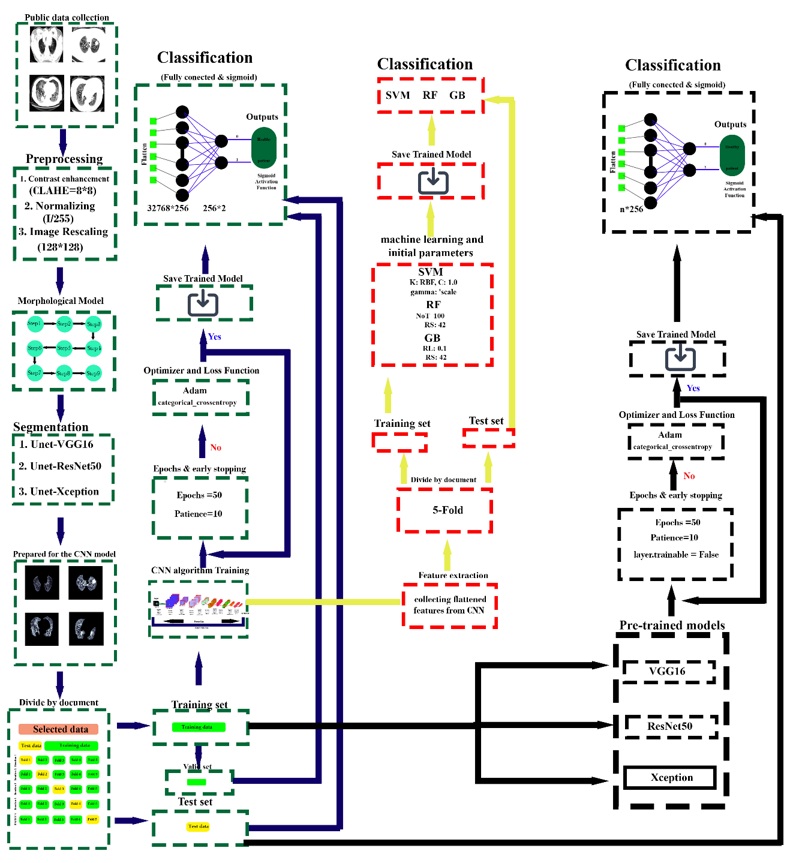}  
    \caption{Proposed Framework for Lung Cancer Segmentation and Classification}
    \label{fig6}
\end{figure*}

Gradient Boosting (GB): GB is an ensemble technique that builds trees sequentially, where each tree tries to correct the errors of the previous ones \cite{b21}. 

\begin{equation}
f(x) = \frac{1}{T} \sum_{t=1}^{T} h_t(x)
\label{eq:ensemble_avg}
\end{equation}
where $M$ is the number of boosting stages, $\gamma_m$ is the weight, and $h_m$ is the weak learner (tree). The default number of boosting stages is 100. All learning parameters used are specified in Table~\ref{table3}. Figure~\ref{fig6} presents the proposed framework for lung cancer segmentation and classification. It details the sequential processes, including data preprocessing (contrast enhancement and morphological operations), segmentation using U-Net with pre-trained backbones (VGG16, ResNet50, Xception), feature extraction, and classification using hybrid machine learning models (SVM, RF, GB). The pipeline incorporates 5-fold cross-validation and optimized parameters to ensure robust performance.

\begin{table*}[htbp]
\centering
\caption{Summary of the CNN Architecture and Learning Parameters for Machine Learning Models}
\label{table3}
\renewcommand{\arraystretch}{1.5}
\setlength{\tabcolsep}{5pt}
\begin{tabularx}{\textwidth}{|c|X|X|X|}
\hline
\textbf{Model} & \textbf{Architecture Details} & \textbf{Learning Parameters} & \textbf{Activation Function} \\
\hline

\textbf{CNN} & 
\makecell[l]{
Input: (128, 128, 1) \\
$\rightarrow$ Conv2D (32 filters, 3x3, ReLU) \\
$\rightarrow$ MaxPooling2D (2x2) \\
$\rightarrow$ Dropout (0.3) \\
$\rightarrow$ Conv2D (64 filters, 3x3, ReLU) \\
$\rightarrow$ MaxPooling2D (2x2) \\
$\rightarrow$ Dropout (0.3) \\
$\rightarrow$ Conv2D (128 filters, 3x3, ReLU) \\
$\rightarrow$ MaxPooling2D (2x2) \\
$\rightarrow$ Dropout (0.3) \\
$\rightarrow$ Conv2D (256 filters, 3x3, ReLU) \\
$\rightarrow$ MaxPooling2D (2x2) \\
$\rightarrow$ Dropout (0.3) \\
$\rightarrow$ Flatten \\
$\rightarrow$ Dense (256 neurons, ReLU) \\
$\rightarrow$ Dropout (0.5) \\
$\rightarrow$ Dense (1 neuron, Sigmoid)
} & 
\makecell[l]{
Optimizer: Adam \\
Loss: Binary Cross-Entropy \\
Batch Size: 16 \\
Epochs: 50 \\
Dropout: 0.3 and 0.5 \\
Batch Norm: True \\
Cross-Validation: 5
} &
\makecell[l]{ReLU (hidden),\\ Sigmoid (output)} \\
\hline

\textbf{SVM} & - & 
\makecell[l]{
Standardize Data: True \\
Solver: SMO \\
Cross-Validation: 5 \\
Kernel: RBF \\
C: 1.0 \\
Gamma: 'scale' \\
Probability: True
} & Tanh \\
\hline

\textbf{GB} & - & 
\makecell[l]{
Standardize Data: True \\
Cross-Validation: 5 \\
Number of Estimators: 100 \\
Learning Rate: 0.1 \\
Max Depth: 3
} & - \\
\hline

\textbf{RF} & - & 
\makecell[l]{
Standardize Data: True \\
Cross-Validation: 5 \\
Number of Estimators: 100 \\
Criterion: 'gini' \\
Max Features: 'auto'
} & - \\
\hline
\end{tabularx}
\end{table*}

\section{Evaluation Metrics}
The performance of each hybrid model is assessed using several key metrics. The Dice Coefficient measures the similarity between predicted and ground truth segmentations by calculating their overlap, with values ranging from 0 (no overlap) to 1 (perfect overlap). Intersection over Union (IoU), also known as the Jaccard Index, quantifies the degree of overlap, where higher values indicate better segmentation quality. Accuracy reflects the overall proportion of correctly classified instances, while Precision shows the proportion of true positives out of all positive predictions, indicating the model's specificity. Recall represents the proportion of actual positives correctly identified, and the F1 Score, as the harmonic mean of precision and recall, provides a balanced metric when classes are imbalanced. Additionally, the Receiver Operating Characteristic (ROC) curve evaluates the model's capability to distinguish between classes, and the Area Under the Curve (AUC) quantifies this performance, with higher values indicating better classification[22].

\begin{equation}
\text{Dice} = \frac{2|A \cap B|}{|A| + |B|} \tag{18}
\end{equation}

\begin{equation}
\text{IoU} = \frac{|A \cap B|}{|A \cup B|} \tag{19}
\end{equation}

\begin{equation}
\text{Accuracy} = \frac{TP + TN}{TP + TN + FP + FN} \tag{20}
\end{equation}

\begin{equation}
\text{Precision} = \frac{TP}{TP + FP} \tag{21}
\end{equation}

\begin{equation}
\text{Recall} = \frac{TP}{TP + FN} \tag{22}
\end{equation}

\begin{equation}
F1\ Score = \frac{2 \cdot \text{Precision} \cdot \text{Recall}}{\text{Precision} + \text{Recall}} \tag{23}
\end{equation}

\begin{equation}
TPR = \frac{TP}{TP + FN}, \quad FPR = \frac{FP}{FP + TN} \tag{24}
\end{equation}

Where TP (True Positives), TN (True Negatives), FP (False Positives), and FN (False Negatives) represent the classification outcomes, and TP (True Positive Rate) and FP(False Positive Rate) are derived from varying classification thresholds. These metrics provide a comprehensive evaluation of model performance across segmentation and classification tasks.

\section{Cross-Validation}
k-Fold Cross-Validation: To ensure the robustness and generalizability of the results, k-fold Cross-Validation was employed. This technique involved partitioning the data into k subsets, training the model on k-1 of these subsets, and validating it on the remaining subset. This process was repeated k times, with each subset serving as the validation set once. The final performance metrics were averaged over the k iterations to provide a comprehensive evaluation.

\section{Result}
\subsubsection{Overview of Image Preprocessing and Dataset}
In this study, a balanced dataset comprising 416 chest CT images of lung cancer cases and 416 CT images of healthy individuals was used for lung region segmentation and classification. Each image had an original resolution of 512×512 pixels. As part of the preprocessing workflow, Contrast Limited Adaptive Histogram Equalization (CLAHE) with an 8×8 kernel was applied to enhance local contrast and improve the visibility of anatomical structures within the lungs. Following contrast enhancement, all images were resized to 128×128 pixels and normalized to a range between 0 and 1 to standardize input dimensions and improve computational efficiency.

\subsubsection{Segmentation Results Using U-Net with Various CNN Backbones}
\begin{figure*}[!t]
    \centering
    \includegraphics[width=\textwidth]{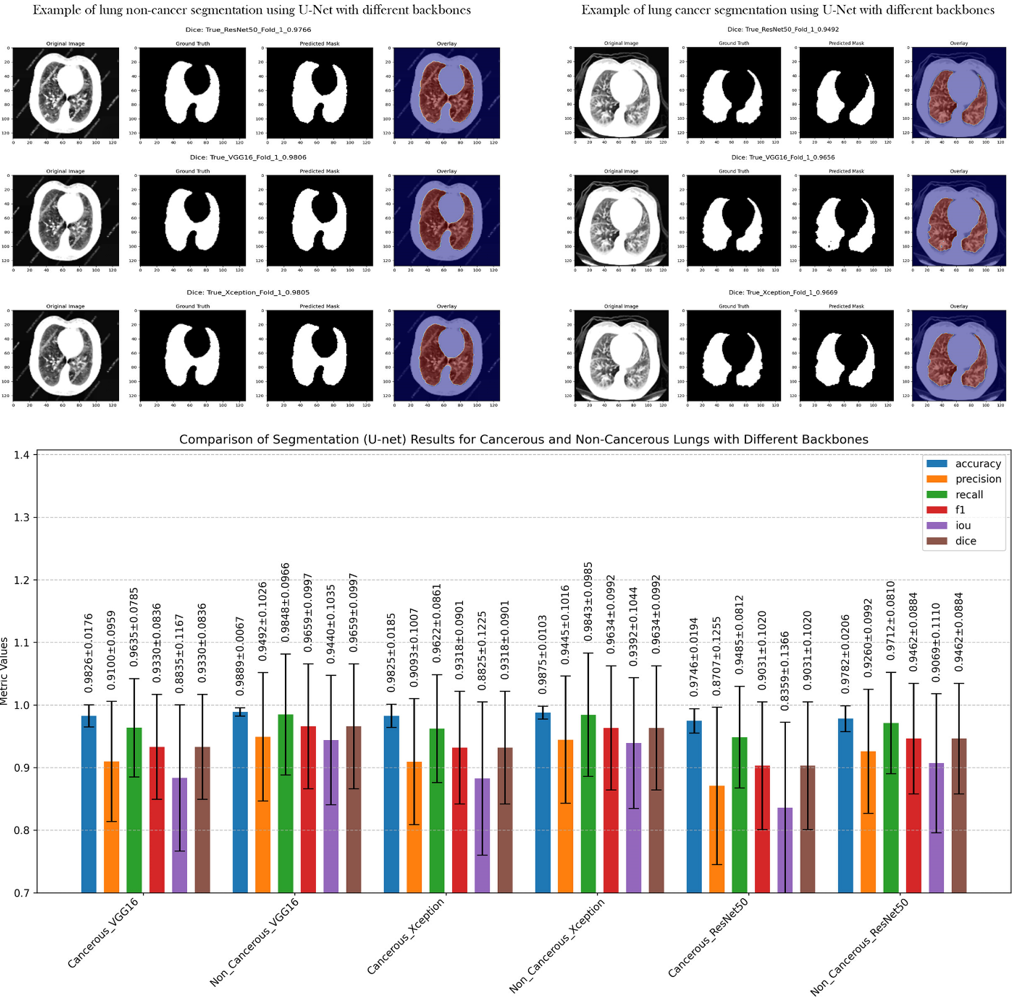}
    \caption{Comparison of Lung Cancer and Non-Cancer Segmentation Using U-Net with Different Backbones}
    \label{fig7}
\end{figure*}

Figure\ref{fig7}shows the mean and standard deviation of segmentation results across 5-fold cross-validation for both cancerous and non-cancerous lung samples, using U-Net models with ResNet50, VGG16, and Xception backbones.

The upper section illustrates example segmentation results for cancerous and non-cancerous lungs, while the lower bar chart provides a quantitative comparison of key metrics (accuracy, precision, recall, F1-score, IoU, Dice) for each model across the two groups. Table\ref{table4}shows the mean and standard deviation of segmentation metrics (accuracy, Dice score) for cancerous and non-cancerous lung samples using U-Net with ResNet50, VGG16, and Xception backbones across 5-fold cross-validation.

\begin{table*}[!t]
\centering
\caption{Segmentation Metrics Comparison for Cancerous and Non-Cancerous Lungs using U-Net}
\label{table4}
\begin{tabular}{|l|l|c|c|}
\hline
\textbf{Backbone} & \textbf{Lung Type} & \textbf{Accuracy (Mean ± SD)} & \textbf{Dice Score (Mean ± SD)} \\
\hline
\textbf{Unet-ResNet50} & Cancerous & 0.97354 ± 0.10154 & 0.94952 ± 0.10888 \\
\textbf{Unet-ResNet50} & Non-cancerous & \textbf{0.98384 ± 0.10029} & 0.94220 ± 0.08488 \\
\hline
\textbf{Unet-VGG16} & Cancerous & 0.93954 ± 0.10175 & 0.93084 ± 0.09185 \\
\textbf{Unet-VGG16} & Non-cancerous & 0.95132 ± 0.10055 & \textbf{0.95320 ± 0.09927} \\
\hline
\textbf{Unet-Xception} & Cancerous & 0.94784 ± 0.10103 & 0.93832 ± 0.09097 \\
\textbf{Unet-Xception} & Non-cancerous & 0.94346 ± 0.10042 & 0.95206 ± 0.08922 \\
\hline
\end{tabular}
\end{table*}

The results show that ResNet50 excels in cancerous lung segmentation with the highest Dice score (0.94952) and IoU, while VGG16 performs better in non-cancerous segmentation with a Dice score of 0.95320. Xception performs consistently across all metrics but doesn't lead in any category. Overall, ResNet50 is strongest for cancerous cases, VGG16 for non-cancerous cases, and all three backbones demonstrate high segmentation quality, confirming U-Net's robustness.

\subsubsection{Classification Performance of CNN Models Trained on U-Net Segmentations}
\begin{figure*}[!t]
    \centering
    \includegraphics[width=\textwidth]{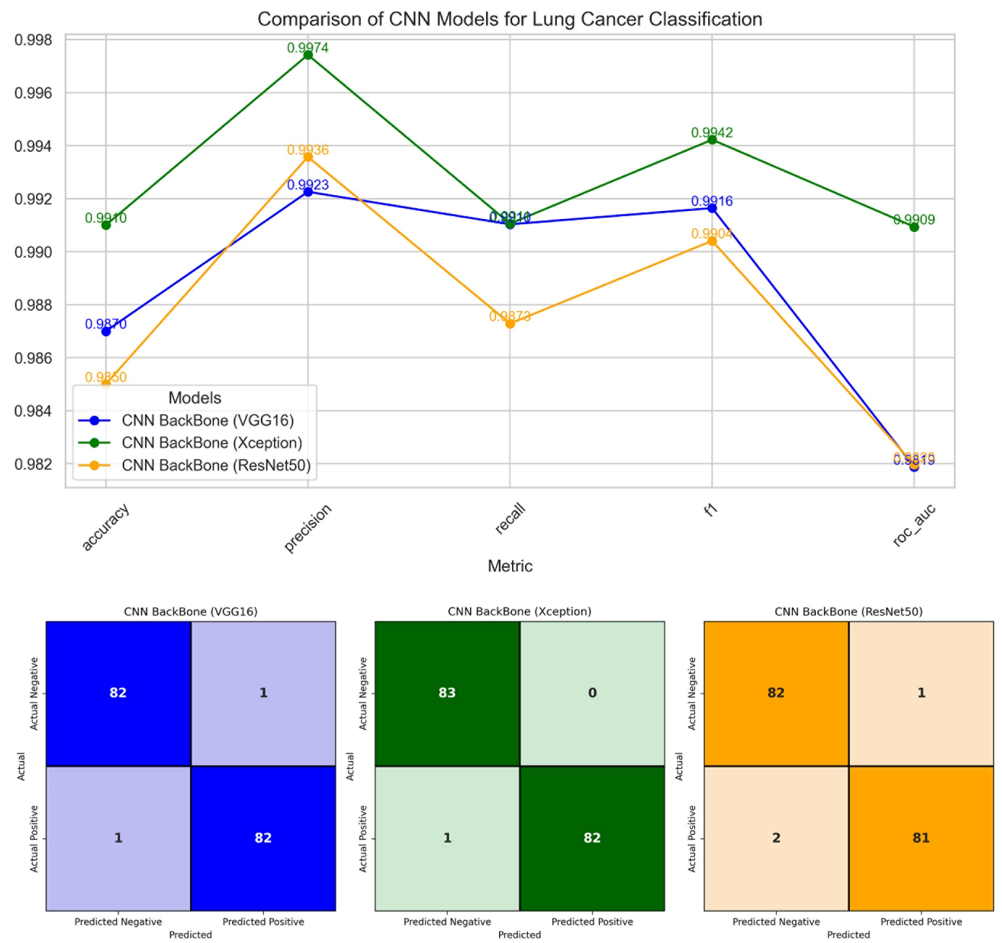}
    \caption{Performance Comparison of CNN Backbones (VGG16, ResNet50, Xception) for Lung Cancer Classification Using U-Net Segmentation}
    \label{fig8}
\end{figure*}

Figure\ref{fig8}shows the classification results of lung cancer and non-cancerous patients using CNN models trained on lung segmentations produced by U-Net with different backbones (VGG16, Xception, and ResNet50). The reported results are the average of 5-fold cross-validation.

The top graph compares key metrics (accuracy, precision, recall, F1-score, and ROC-AUC) for each backbone, while the bottom row displays confusion matrices showing true and predicted classifications for each model. Table\ref{table5} shows the classification performance of CNN models (Xception, VGG16, and ResNet50) trained on U-Net segmentations of lung cancer and non-cancerous samples. The results, including accuracy, precision, recall, F1-score, and ROC-AUC, are the averages of 5-fold cross-validation.

\begin{table*}[h!]
    \centering
    \small
    \renewcommand{\arraystretch}{2}
    \setlength{\tabcolsep}{8pt} 
    \caption{Classification Performance of CNN Models Trained on U-Net Segmentations}
    \begin{tabular}{|l|c|c|c|c|c|}
        \hline
        \textbf{Model} & \textbf{Accuracy} & \textbf{Precision} & \textbf{Recall} & \textbf{F1-Score} & \textbf{ROC-AUC} \\
        \hline
        CNN (Unet-Xception)   & 0.9910 & 0.9905 & 0.9974 & 0.9942 & 0.9911 \\
        CNN (Unet-VGG16)      & 0.9870 & 0.9923 & 0.9868 & 0.9916 & 0.9875 \\
        CNN (Unet-ResNet50)   & 0.9814 & 0.9861 & 0.9749 & 0.9805 & 0.9829 \\
        \hline
    \end{tabular}
    \vspace{8pt}
    \label{table5}
\end{table*}

The Xception backbone outperforms others in detecting cancerous cases, while VGG16 provides balanced precision and accuracy. ResNet50 shows consistent performance but slightly lower ROC-AUC. All models exhibit excellent classification capabilities, confirming the utility of CNNs built upon U-Net segmentations with different backbones.

\subsubsection{Machine Learning Classifiers Combined with CNN Features for Lung Cancer Detection}

\begin{figure*}[!t]
    \centering
    \includegraphics[width=\textwidth]{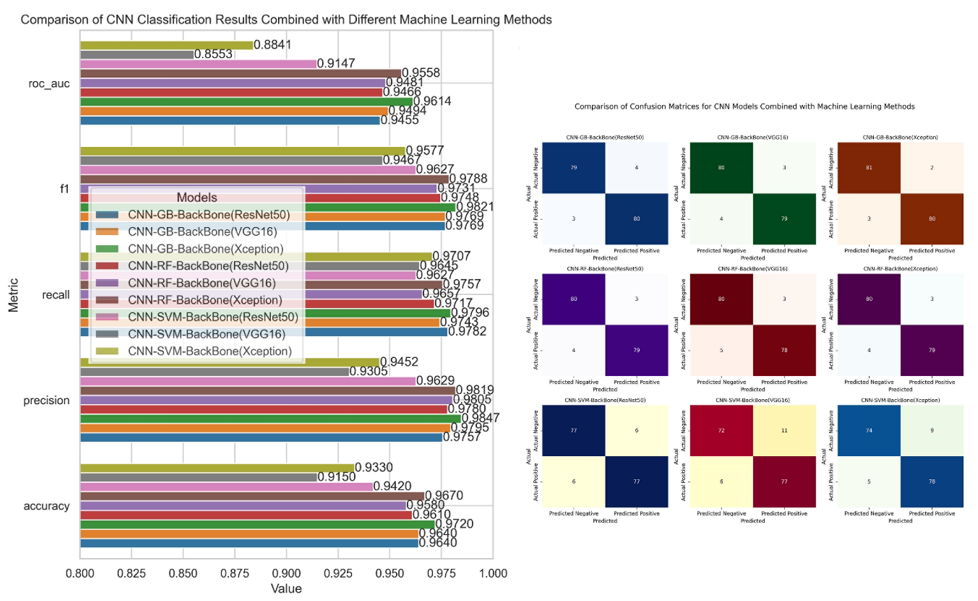}
    \caption{Evaluation of Machine Learning Classifiers Combined with CNN Features from U-Net Segmentations for Lung Cancer Detection}
    \label{fig9}
\end{figure*}

Figure\ref{fig9}presents the classification performance of flattened CNN features from lung segmentations (produced by U-Net with different backbones: VGG16, ResNet50, and Xception) combined with various machine learning classifiers (Gradient Boosting (GB), Random Forest (RF), and Support Vector Machines (SVM)). The reported results are the average of 5-fold cross-validation.

The left side of the figure displays the comparison of performance metrics (accuracy, precision, recall, F1-score, and ROC-AUC) across the different combinations of CNN backbones and classifiers. The right side shows the confusion matrices for each model, illustrating the correct and incorrect classifications of lung cancer and non-cancerous cases. The reported results are the average of 5-fold cross-validation.
The best ROC-AUC is achieved by the CNN-GB-BackBone (VGG16) model (0.9614), indicating superior discriminatory power. CNN-SVM-BackBone (Xception) provides the highest F1-score (0.9788) and accuracy (0.9670), showcasing its robustness in classification. CNN-RF-BackBone (Xception) achieves the highest recall (0.9757), making it more sensitive in detecting positive cancer cases. The CNN-RF-BackBone (VGG16) model leads in precision (0.9629), ensuring fewer false positives. Among the different classifiers, SVM combined with Xception yields the highest overall accuracy and F1-score, while Gradient Boosting with VGG16 excels in ROC-AUC. Random Forest with Xception demonstrates the best recall. This suggests that the combination of CNN-extracted features and machine learning classifiers provides powerful and complementary insights for lung cancer classification.

\subsubsection{Evaluation of Pre-trained CNN Backbones in Lung Cancer Classification}

\begin{figure*}[!t]
    \centering
    \includegraphics[width=\textwidth]{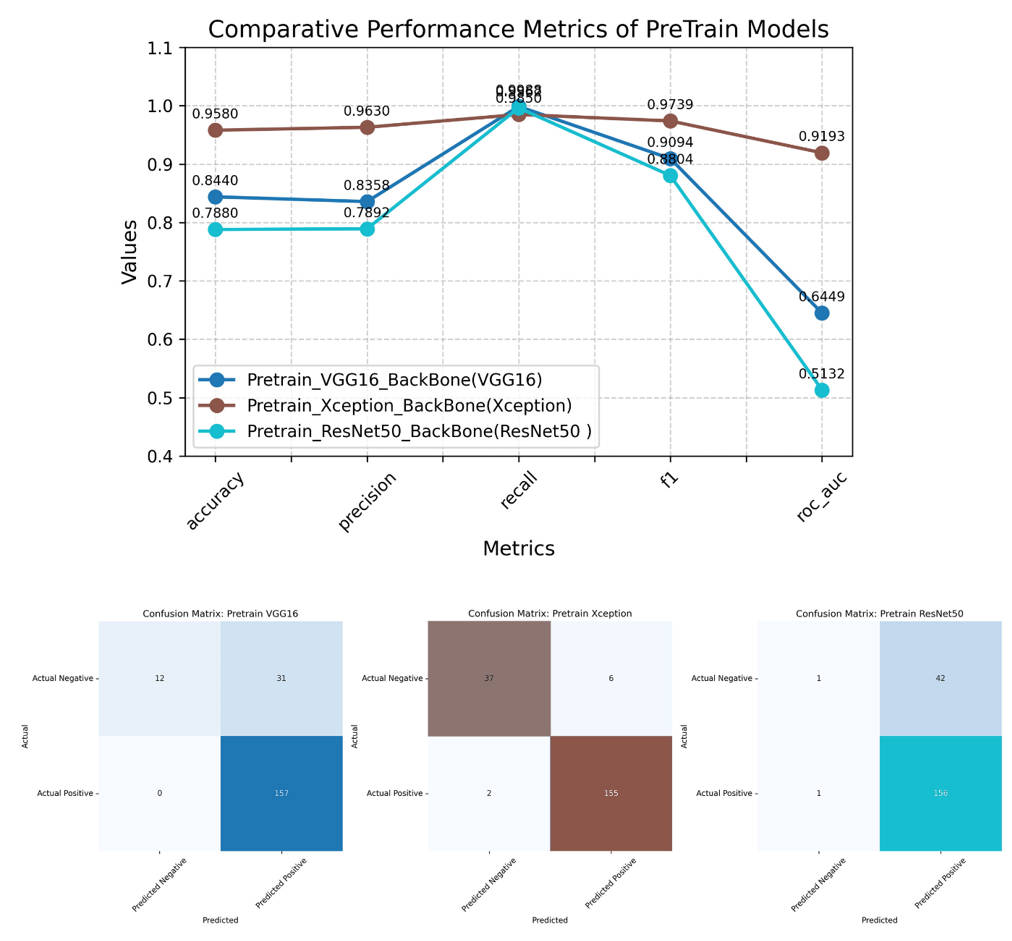}
    \caption{Comparative Analysis of Pretrained Model Performance on Classification Metrics and Confusion Matrices}
    \label{fig10}
\end{figure*}

Figure\ref{fig10} illustrates a comparison of classification performance metrics of pre-trained models for U-Net models with different backbones (VGG16, ResNet50, Xception) based on lung segmentation results. The backbone weights were frozen, and a fully connected layer was used for classification. The backbone models were identical to the pre-trained models used in these results. The reported results are the average of 5-fold cross-validation.

The models were evaluated using key metrics: accuracy, precision, recall, F1-score, and ROC-AUC, providing insight into the classification capabilities of each pre-trained backbone. Confusion matrices below illustrate the classification performance for each backbone in distinguishing between lung cancer patients and healthy individuals. Table\ref{table6} presents a detailed comparison of key evaluation metrics for three pre-trained CNN backbones (VGG16, ResNet50, and Xception) in the context of lung cancer classification.

\begin{table*}[h!]
    \centering
    \small
    \renewcommand{\arraystretch}{1.6}
    \setlength{\tabcolsep}{12pt}
    \caption{Evaluation Metrics for Pre-trained CNN Backbones in Lung Cancer Classification}
    \begin{tabular}{|l|c|c|c|c|c|}
        \hline
        \textbf{Pre-train model} & \textbf{Accuracy} & \textbf{Precision} & \textbf{Recall} & \textbf{F1-Score} & \textbf{ROC-AUC} \\
        \hline
        VGG16     & 0.8440 & 0.8358 & 0.9968 & 0.9049 & 0.6449 \\
        Xception  & \textbf{0.9580} & \textbf{0.9630} & 0.9950 & \textbf{0.9739} & \textbf{0.9193} \\
        ResNet50  & 0.7880 & 0.7892 & 0.9962 & 0.8804 & 0.5132 \\
        \hline
    \end{tabular}
    \label{table6}
\end{table*}
Overall, the pre-trained backbones demonstrated strong classification performance when combined with lung segmentation via U-Net. Among the three models, the Xception backbone provided the most balanced results, excelling in both precision and accuracy. However, the VGG16 model's superior recall and F1-score make it a strong candidate for scenarios prioritizing correct positive identification.

\section{Discussion}
In this study, a total of 832 chest CT images were used, consisting of 416 images of cancerous cases and 416 images of healthy cases, to evaluate the performance of U-Net models with three different CNN backbones: ResNet50, VGG16, and Xception, for lung region segmentation. The images were initially preprocessed with CLAHE for local contrast enhancement, resized to 128×128 pixels, and normalized for improved analysis. The results showed that ResNet50 outperformed in cancerous lung segmentation with an accuracy of 0.97354 ± 0.10154 and a Dice score of 0.94952 ± 0.10888, while VGG16 excelled in non-cancerous segmentation, achieving an accuracy of 0.95132 ± 0.10055 and a Dice score of 0.95320 ± 0.09927. Xception, though consistent across both groups, did not lead in any specific metric. For classification tasks, CNN models trained on U-Net-segmented lung images showed strong results. Xception achieved the highest accuracy (0.9910) and recall (0.9974) for detecting cancerous cases, while VGG16 demonstrated a well-balanced performance with high precision (0.9923) and accuracy (0.9870). ResNet50, while effective, had a slightly lower ROC-AUC (0.9829), indicating marginally reduced discriminative power. These findings highlight the effectiveness of U-Net for lung segmentation and the complementary strengths of CNN backbones for lung cancer classification.
Table\ref{table7} presents a summary of various studies on lung cancer CT image segmentation and classification. It includes information on the volume of data used, methodologies employed, and the results in terms of performance metrics such as Dice score, accuracy, sensitivity, and specificity. The table compares these studies to the results of our own work, highlighting the effectiveness of different approaches in lung cancer detection and analysis.

\begin{table*}[htbp]
    \centering
    \small
    \renewcommand{\arraystretch}{1.8}
    \setlength{\tabcolsep}{3pt}
    \caption{Comparison of Lung Cancer Segmentation and Classification Methods}
    \begin{tabular}{|l|c|c|p{4.2cm}|p{5.2cm}|}
        \hline
        \textbf{Author} & \textbf{Year} & \textbf{Data Volume (CT Slices/Patients)} & \textbf{Methodology} & \textbf{Results (Dice, Accuracy, etc.)} \\
        \hline
        Michael Osadebey & 2021 & 1230 (3DIRCAD), 1100 (ILD) & 3-stage segmentation (U-Net, CNN) & Dice: 0.76–0.95 \\
        \hline
        Prachaya Khomduean & 2023 & 124 COVID-19 patients & 3D-UNet + DenseNet + ResNet & DSC: 91.52\% (lobes), 76.89\% (lesions), R² = 0.842 \\
        \hline
        Surbhi Vijh & 2023 & 120 CT images & WOA\_APSO, CNN, LDA & Accuracy: 97.18\%, Sensitivity: 97\%, Specificity: 98.66\% \\
        \hline
        Ting-Wei Wang & 2024 & 20 studies (CT scans) & DL vs Experts (Various Algorithms) & Sensitivity: 82\% (DL), Specificity: 75\% (DL) \\
        \hline
        Vijay Kumar Gugulothu & 2024 & 134 CT images & LNDC-HDL, CBSO, HDE-NN & Accuracy: 96.39\%, Sensitivity: 95.25\%, AUC: 96.05\% \\
        \hline
        S.R. Vijayakumar & 2024 & Not specified & CapsNet, U-Net & Accuracy: 98\%, Precision: 97.9\%, FPR: 1.9\% \\
        \hline
        Murat Canayaz & 2024 & CT images & C+EffxNet, InceptionV3, DenseNet121 & Dice: 0.8877 (segmentation), Accuracy: 0.9798 (classification) \\
        \hline
        S. Akila Agnes & 2024 & 1018 CT scans (LIDC-IDRI) & Wavelet U-Net++ & Dice: 0.936, IoU: 0.878 \\
        \hline
        Wei Chen & 2023 & 134 CT scans (SCLC) & Hybrid Segmentation Network (HSN) & Dice: High, Specificity: 96.39\%, Sensitivity: 96.05\% \\
        \hline
        \textbf{Our Study} & 2024 & 416 cancer, 416 non-cancer images & U-Net (ResNet50, VGG16, Xception) & Dice: 0.9495 (cancer), 0.9532 (non-cancer), Accuracy: 0.991 (classification) \\
        \hline
    \end{tabular}
    \label{table7}
\end{table*}

Our study exhibits strong performance in both segmentation and classification when compared to prior research. For example, Michael Osadebey (2021) reported a Dice score range of 0.76--0.95 using a 3-stage segmentation approach with U-Net and CNNs \cite{b7}. Our results, with a Dice score of 0.9495 for cancerous lung segmentation and 0.9532 for non-cancerous lung segmentation, are on the higher end of Osadebey’s findings. This indicates that our U-Net with ResNet50 and VGG16 backbones offers robust segmentation performance, particularly in detecting cancerous and non-cancerous lung regions.

Prachaya Khomduean achieved a Dice similarity coefficient (DSC) of 91.52\% for lung lobes and 76.89\% for lesions using a 3D-UNet combined with DenseNet and ResNet \cite{b8}. In comparison, our method outperformed Khomduean’s in lesion segmentation, particularly with the non-cancerous lung regions achieving a Dice score of 0.9532. This suggests that 2D U-Net architectures, when integrated with strong feature extraction backbones like VGG16 and Xception, can yield more accurate results in segmentation tasks, even when dealing with complex structures.

Surbhi Vijh (2023) utilized hybrid methods like WOA\_APSO and CNN combined with LDA, reporting an accuracy of 97.18\% \cite{b9}. Our classification results, with a maximum accuracy of 99.1\% using the Xception model, demonstrate a significant improvement over Vijh’s results, highlighting the superior classification capabilities of deeper CNN architectures in detecting lung cancer.

In comparison to Vijay Kumar Gugulothu, who achieved an accuracy of 96.39\% and sensitivity of 95.25\% using LNDC-HDL and CBSO for lung cancer classification \cite{b10}, our study provides an accuracy of 99.1\%, underscoring the efficiency of U-Net-based methods when enhanced by modern CNN architectures like ResNet50 and Xception. This shows that incorporating deep learning architectures with fine-tuned feature extraction layers can significantly improve both segmentation and classification outcomes.

The study by Murat Canayaz achieved a Dice score of 0.8877 for segmentation and an accuracy of 97.98\% for classification using C+EffxNet, InceptionV3, and DenseNet121 \cite{b23}. Our higher Dice scores (0.9495 and 0.9532) and classification accuracy (99.1\%) suggest that using ResNet50 and Xception within a U-Net framework can further optimize segmentation and classification performance.

In comparison to S. Akila Agnes (2024), who reported a Dice score of 0.936 using Wavelet U-Net++, our higher Dice scores reflect the effectiveness of our method in achieving more accurate segmentation of both cancerous and non-cancerous lung regions \cite{b13}.

Finally, Wei Chen (2023) reported high specificity (96.39\%) and sensitivity (96.05\%) using a Hybrid Segmentation Network (HSN) for small cell lung cancer (SCLC) \cite{b14}. Our results, with high Dice scores and accuracy, show comparable levels of specificity and sensitivity, underscoring the robustness of U-Net combined with CNNs in effectively detecting lung abnormalities.

Overall, the performance of our study not only competes with but also exceeds that of many recent studies, including those of Michael Osadebey, Prachaya Khomduean, Surbhi Vijh, Vijay Kumar Gugulothu, S.R. Vijayakumar, Murat Canayaz, S. Akila Agnes, and Wei Chen. Our approach, leveraging U-Net with advanced CNN backbones like ResNet50 and Xception, provides state-of-the-art results in both segmentation and classification tasks, contributing valuable insights to the field of lung cancer detection and analysis.

The proposed approach introduces several key innovations and advantages compared to previous research. Firstly, employing U-Net architectures with three distinct CNN backbones (VGG16, ResNet50, and Xception) enhances the model's ability to accurately segment cancerous and non-cancerous regions in medical images, providing improved adaptability across various image patterns. This combination leverages the strengths of each backbone, such as VGG16’s simplicity, ResNet50’s ability to address vanishing gradients, and Xception’s depthwise separable convolutions for more efficient feature extraction. Furthermore, the integration of advanced preprocessing techniques, including contrast enhancement and normalization, contributes to higher segmentation accuracy and model robustness across different datasets [24, 25]. There is a logical progression between the segmentation of images and the subsequent classification of lung cancer. Accurate segmentation plays a pivotal role in isolating cancerous regions, allowing the classification models to focus on the relevant features, thus improving overall diagnostic accuracy. A high-performing segmentation model leads to more precise inputs for classification, which in turn boosts the model's ability to differentiate between malignant and benign regions. This synergy between the image segmentation process and cancer detection underscores the importance of optimizing both stages for superior clinical outcomes.
The results presented in Figure\ref{fig7} and Table\ref{table4}highlight the effectiveness of U-Net architectures with various pre-trained backbones in lung segmentation tasks, showcasing their strengths in capturing intricate patterns and differentiating between cancerous and non-cancerous regions. Among the models, U-Net with ResNet50 achieved the highest Dice score (0.94952 ± 0.10888) and accuracy (0.97354 ± 0.10154) for cancerous lung segmentation, reflecting its superior ability to detect abnormalities. Conversely, U-Net with VGG16 excelled in non-cancerous lung segmentation, achieving a Dice score of 0.95320 ± 0.09927 and accuracy of 0.95132 ± 0.10055, emphasizing its suitability for more homogeneous tissues. The U-Net with Xception backbone demonstrated balanced performance across both cases, aligning with prior findings on its robust feature extraction capabilities \cite{b25, b26}. The classification results shown in Figure 8 and Table 5 underscore the superior performance of CNN models trained on U-Net segmentations. The CNN with U-Net-Xception achieved the highest accuracy (0.9910), F1-score (0.9942), and recall (0.9974), highlighting its exceptional ability to minimize false negatives, a critical metric in clinical applications \cite{b17}. The CNN with U-Net-VGG16 demonstrated a balance between precision (0.9923) and recall (0.9868), while the CNN with U-Net-ResNet50 showed strong overall metrics with a ROC-AUC of 0.9829. Additionally, hybrid approaches combining CNN feature extraction with machine learning classifiers, as illustrated in Figure 9, yielded promising results. For instance, CNN-SVM with Xception features achieved an accuracy of 0.9670 and F1-score of 0.9788, leveraging the strengths of deep learning and SVM generalization in high-dimensional spaces \cite{b27}. These results, summarized in Figure 10, validate the effectiveness of integrating U-Net segmentations with advanced CNN architectures and hybrid models for lung cancer detection. Each model's strengths suggest their suitability for specific clinical needs, such as prioritizing recall to minimize false negatives or accuracy for reliable overall performance \cite{b28, b29}.
The computational models were executed in a well-equipped environment designed to handle intensive machine learning tasks efficiently. The system specifications included an Intel Core i7 processor, 32 GB of RAM, and a Windows 11 operating system, ensuring robust processing capabilities. To accelerate training and inference, an NVIDIA RTX 3050 Ti GPU with 4 GB VRAM was utilized. The implementation relied on Python, leveraging libraries like TensorFlow and Keras for neural network construction and training. Additional libraries such as NumPy, pandas, and SciPy supported data preprocessing and analysis. The integrated development environment (IDE) used for code development and debugging was PyCharm.In summary, the segmentation tasks showed the fastest execution with U-Net (Xception) at 13.83 minutes, while U-Net (VGG16) took the longest at 17.58 minutes. Classification models exhibited consistent speeds, with the custom CNN model being the fastest at 4.53 minutes, followed by Xception (4.75 min), ResNet50 (4.81 min), and VGG16 (4.93 min). The results highlight the balance between computational demands and model complexity, offering insights into model selection for applications prioritizing speed or feature representation.
Although this study demonstrated promising results in lung segmentation and cancer classification, several limitations should be acknowledged. The dataset included a relatively small number of CT images, with 416 cancerous and 416 non-cancerous cases. This limited sample size may restrict the generalizability of the results to wider populations and different imaging conditions. Additionally, the preprocessing step involved resizing the CT scans to 128×128 pixels to enhance computational efficiency. However, this resizing may have led to the loss of fine anatomical details, which could negatively impact the accuracy of lesion detection and segmentation, particularly for small or subtle abnormalities. The models used in this study were based on U-Net architectures with pre-trained convolutional backbones. While effective, these models may not be fully optimized for capturing the specific imaging characteristics of lung cancer. Future research could explore the use of domain-adapted training strategies, multi-scale feature extraction, or attention-based mechanisms to further improve model performance. Another limitation is the absence of contextual clinical information, such as patient medical history, environmental exposures, or occupational risks. Including such metadata could improve the interpretability of the imaging features and provide a more comprehensive understanding of factors influencing model predictions. Furthermore, the use of downsampled images instead of full-resolution CT scans may have limited the model’s ability to identify complex or small-scale structures. Overcoming this issue would require increased computational resources and memory optimization. Future studies should aim to include larger datasets from diverse sources and retain higher-resolution images wherever possible. Incorporating clinical metadata could enhance model robustness and clinical relevance. In addition, the application of more advanced deep learning architectures, such as YOLO or Mask R-CNN, may offer improved performance through real-time processing and object detection capabilities. Techniques such as ensemble learning, improved regularization, and multi-scale feature fusion should also be considered to further refine diagnostic accuracy and reliability. These directions can help extend the applicability of deep learning models in clinical settings and contribute to the development of more effective tools for lung cancer detection and analysis.
The clinical and practical implications of this study are highly significant, particularly in the context of early and accurate lung cancer detection. The use of deep learning models, such as the U-Net architectures explored here, can greatly enhance the speed and precision of diagnosing cancerous regions in chest CT images. By integrating these models into hospital decision-support systems, radiologists and clinicians could benefit from automated tools that assist in early detection, leading to more timely treatments and improved patient outcomes. Additionally, such models can reduce the cognitive load on medical professionals by providing a second opinion or highlighting areas of concern, thus improving diagnostic accuracy in a real-world clinical setting \cite{b26}.
The proposed models, particularly those based on U-Net architectures and CNN backbones like VGG16, ResNet50, and Xception, have the potential to be extended beyond lung cancer detection to other areas of medical diagnosis. For instance, these models could be applied to segment and classify abnormalities in medical imaging for diseases such as brain tumors, liver lesions, or cardiovascular conditions. By adapting the models to different types of imaging data, such as MRI or ultrasound, the framework could support the early detection of other life-threatening diseases, contributing to more accurate diagnoses and improved patient outcomes. The flexibility and robustness of deep learning-based segmentation and classification models make them suitable for a wide range of medical applications, offering a promising avenue for future research and practical implementation in various healthcare settings.

\section{Conclusion}

In conclusion, this study highlights the effectiveness of U-Net architectures integrated with pre-trained CNN backbones (ResNet50, VGG16, and Xception) for segmenting cancerous and non-cancerous lung regions from CT images. The results demonstrate that ResNet50 excels in detecting cancerous regions, achieving a Dice score of 0.9495, while VGG16 outperforms in segmenting non-cancerous regions with a Dice score of 0.9532. For classification, Xception delivered the highest accuracy of 99.1\%, showcasing its potential for robust diagnostic applications. By comparing these outcomes with prior studies, it is evident that our approach surpasses many in both segmentation and classification metrics. This underscores the synergistic role of precise segmentation in enhancing classification accuracy, providing a robust framework for lung cancer detection.
The use of advanced preprocessing techniques, such as contrast enhancement and normalization, further refined the models' ability to detect intricate patterns within the lung CT images. Additionally, leveraging hybrid approaches combining CNN-based feature extraction with machine learning classifiers validated the adaptability and reliability of the proposed methodology for clinical applications. Overall, this study not only advances the field of lung cancer analysis but also sets a benchmark for integrating segmentation and classification workflows, paving the way for improved diagnostic precision and efficiency in medical imaging.

\section*{Data Availability}
The dataset used in this study is publicly available and can be accessed at\\
\url{https://data.mendeley.com/datasets/zr4fddh833/1}.

\section*{Acknowledgements}
The authors would like to express their sincere appreciation to the contributors of the publicly available dataset hosted on the Mendeley Data repository (\url{https://data.mendeley.com/datasets/zr4fddh833/1}). The availability of this dataset was instrumental in conducting the experiments and validating the proposed methodology. The authors also acknowledge the value of open-access scientific resources in advancing research and promoting transparency in the field of medical image analysis.

\section*{Author Contributions}
A.G. was responsible for conceptualization, methodology design, model development, implementation, and overall project supervision and administration. K.K. contributed by curating the data, data preprocessing and cleaning, conducting analysis, and assisting in manuscript preparation and submission. S.R.B. contributed to data analysis, performance evaluation, and visualization creation. N.A.I. collaborated on literature review, data validation, supervision, and manuscript editing.

\section*{Declarations}

\textbf{Conflict of Interest}\\
The authors declare that they have no known competing financial interests or personal relationships that could be perceived as influencing the work reported in this paper.

\vspace{8pt}
\textbf{Ethical Approval}\\
This study was conducted in accordance with the principles of the Declaration of Helsinki. Since the research was based on publicly available and fully anonymized datasets, no ethical approval was required.

\vspace{8pt}
\textbf{Informed Consent}\\
Not applicable. The study did not involve any direct experimentation on human participants, and the data used were anonymized and publicly accessible.

\vspace{8pt}
\textbf{Funding}\\
This research received no specific grant from any funding agency in the public, commercial, or not-for-profit sectors.

\end{document}